\documentclass[preprint,floats,floatfix,nofootinbib]{revtex4}

\usepackage{amsbsy}
\usepackage{amssymb}
\usepackage{amsmath}
\input{epsf}
\usepackage{graphicx}
\usepackage{color}

\usepackage{amssymb}
\usepackage{bm}
\def\spose#1{\hbox to 0pt{#1\hss}}

\def\lta{\mathrel{\spose{\lower 3pt\hbox{$\mathchar"218$}}
     \raise 2.0pt\hbox{$\mathchar"13C$}}}
\def\gta{\mathrel{\spose{\lower 3pt\hbox{$\mathchar"218$}}
     \raise 2.0pt\hbox{$\mathchar"13E$}}}

\def\beq{\begin{equation}}
\def\eeq{\end{equation}}
\def\bea{\begin{eqnarray}}
\def\eea{\end{eqnarray}}

\def\x{{\rm x}}

\def\n{{\rm n}}
\def\p{{\rm p}}
\def\e{{\rm e}}

\begin{document}
\title{The sum of Love: Exploring the effective tidal deformability of neutron stars}


\author{N. Andersson}
\affiliation{Mathematical Sciences and STAG Research Centre, University of Southampton, Southampton 
SO17 1BJ, UK}

\author{P. Pnigouras}
\affiliation{Mathematical Sciences and STAG Research Centre, University of Southampton, Southampton 
SO17 1BJ, UK}


\begin{abstract}
Finite size effects come into play during the late stages of neutron star binary inspiral, with the tidal deformability of the supranuclear density matter leaving an imprint on the gravitational-wave signal. As  demonstrated in the case of  GW170817, this leads to a constraint on the neutron star radius (and hence the equation of state). A deeper understanding of the tidal response requires an analysis of both the state and composition of matter. While these aspects may not have dramatic impact, they could lead to systematic effects that need to be kept in mind as the observational data become more precise. As a step in this direction we explore the role of the  composition of matter, which is likely to remain ``frozen'' during the late stages of binary inspiral. We provide the first in-depth analysis of the problem, including estimates of how  composition impacts on the effective tidal deformability. The results provide improved insight into how aspects of physics that tend to be ``ignored'' impact on binary neutron star gravitational-wave signals.
\end{abstract}

\maketitle

\section{Introduction}

The breakthrough detection of signals from inspiralling and merging black-hole binaries \cite{gw150914,catalog} and the spectacular neutron star merger event GW170817 \cite{gw170817} demonstrate the promise of gravitational-wave astronomy. As the sensitivity of the detectors improves  we can expect other exciting discoveries. In particular, further events involving neutron stars should allow us to put tighter constraints on the physics of matter under extreme conditions.

A typical neutron star binary system may spend as much as 15 minutes in the sensitivity band of advanced ground-based interferometers (above 10 Hz). The detection of, and extraction of parameters from, such signals is of great importance for both astrophysics and nuclear physics. From the astrophysics point of view observed event rates should lead to insight into the formation channel(s) for these systems, while the nuclear physics aspects relate to the poorly constrained equation of state for matter at supranuclear densities (see for example \cite{LOFT}). 
Focussing on the nuclear physics, binary neutron star signals constrain the equation of state of supranuclear matter in two ways. First of all, finite size effects  impact on the inspiral signal. The deformability of the stellar fluid leaves an  imprint on the late-time chirp, an effect encoded in the tidal Love numbers \cite{hind1,hind2,hind3,agat}. Meanwhile, oscillations of the merger remnant, which depend on the hot equation of state, are expected to leave a robust signature \cite{baus1,baus2,baus3,bern,luc1,luc2}. However, the high-frequency nature of the merger signal makes it difficult to detect with the current generation of detectors (even at design sensitivity) \cite{clark,ligo_pm}.    

In this paper we focus on the tidal deformability. Our aim is simple; we want to explore to what extent the composition of the neutron star matter enters the problem. The motivation for this is clear. As the star is deformed by the tidal interaction, matter is  driven out of equilibrium and it is easy to argue that the  the relevant nuclear reactions are too slow to re-establish equilibrium on the timescale of inspiral\footnote{Note that the situation is different for the tide in an ordinary stellar binary, where the orbital evolution is excruciatingly slow. In that context, the assumption of a ``static'' tide, encoding the deformability in the Love numbers, does not need to be questioned.}. This is clear from, for example, the estimates in \cite{haensel}. At the simplest level, the relevant equilibration timescales are
\beq
t_M \sim {2\ \mathrm{months} \over T_9^6} \ , \quad t_D \sim {20\ \mathrm{ s} \over T_9^4}\ , 
\eeq
for the modified and direct Urca reaction, respectively. The temperature is scaled to hot systems, $T_9=T/10^9$~K, but inspiralling neutron stars are old and   cold, typically in the range $T_9\le 0.01$, which would make both $t_M$ and $t_D$ much longer than the time it takes a given system to move through the sensitivity band of a ground-based interferometer (minutes). Moreover, at the expected temperature the star's interior should be superfluid, in which case reactions are exponentially suppressed. However,  superfluidity brings additional aspects to the problem (which warrant a separate discussion) so we will focus on the (simpler) case of non-superfluid matter in the following. 

The upshot of the equilibration argument is that  the equation of state is no longer barotropic, as has been assumed in virtually every previous analysis of the tidal problem (most of which follow the steps laid out in \cite{hind2,hind3}). Instead, we want to establish to what extent a frozen matter composition leads to a noticeable effect on, for example, the Love numbers\footnote{The issue also arises in connection with the so-called I-Love-Q relations \cite{ilq}, but we will not consider that aspect here.}, and if this in turn impacts on the extraction of neutron star parameters from an observed signal. The effect may well be small enough that we can ignore it, given the anticipated observational ``errors'', but we need to make sure that this is the case. In essence, we want to quantify the systematic ``errors'' associated with the current models. 

\section{The adiabatic  problem}

In order to set the stage for the discussion, it is useful to remind ourselves of the context.  The tide raised by a binary companion (here treated as a point particle, which should be a good enough approximation for our purposes) induces a linear response in the primary. 
In order to quantify this response we solve the linearised fluid equations in Newtonian gravity. Assuming that the star is non-rotating, we first of all have the  perturbed continuity equation 
\beq
\partial_t \left( \Delta \rho + \rho \nabla_i  \xi^i \right)  = 0 
\label{continuity}
\eeq
where $\xi^i$ is the  displacement vector associated with the Lagrangian perturbation \cite{fs78a}
\beq
\Delta  = \delta + \mathcal L_\xi
\eeq
(with $\delta$ the corresponding Eulerian variation and $\mathcal L_\xi$ the Lie derivative along $\xi^i$) such that the perturbed velocity is given by  
\beq
\Delta v^i = 
\delta  v^i = \partial_t  \xi^i
\label{delv}
\eeq
The perturbed Euler equation is
\beq
\partial_t^2  \xi_i + {1\over \rho}  \nabla_i \Delta p - {\Delta \rho\over \rho^2}  \nabla_i p + \nabla_i \Delta \Phi = - \nabla_i \chi
\label{eulp}
\eeq
or, equivalently, provided that the background is in hydrostatic equilibrium,
\beq
\partial_t^2  \xi_i +{1\over \rho}  \nabla_i \delta p - {1\over \rho^2} \delta \rho \nabla_i p + \nabla_i \delta \Phi = - \nabla_i \chi
\label{eul2}
\eeq
(where it is useful to keep in mind that the Lagrangian variation commutes with the gradient $\nabla_i$).

We also  have the Poisson equation for the gravitational potential 
\beq
\nabla^2 \delta \Phi = 4\pi G \delta \rho
\label{pois1}
\eeq
while the tidal potential due to the presence of the binary partner (which generates the fluid perturbation) is given by a solution to $\nabla^2 \chi = 0$.

 In  a coordinate system centred on the primary, which we will take to have mass $M_\star$, we have \cite{holai}
\beq
\chi = - {GM' \over |\boldsymbol r - \boldsymbol D (t)|}= - GM'\sum_{l\ge2} \sum_{m=-l}^l {W_{lm} r^l \over D^{l+1}(t) } Y_{lm} e^{-im\psi(t)}
\label{tidpot}
\eeq
where $M'$ is the mass of the secondary. The orbit of the companion is taken to be in the plane $[D(t), \pi/2, \psi(t)]$ where $D$ is the binary separation and $\psi$ is the orbital phase. For $l=2$ (which leads to the main contribution to the gravitational-wave signal) we have  \cite{pt77}
\beq
W_{20}= - \sqrt{\pi/5} \label{W20} \ , \quad 
W_{2\pm2} = \sqrt{3\pi/10} \ , \quad
W_{2\pm1} = 0 \ .
\eeq
The last result follows from symmetry; $W_{lm}$ must vanish for all odd $l+m$.

We need to work out the deformation of the star due to the companion's tidal field. That is, we are looking for a solution to \eqref{eul2} given the specific form for $\chi$. 
The  problem is usually explored in the frequency domain. Simplistically, this is often taken to mean that the time-dependence of perturbed quantities is harmonic, proportional to $e^{i\omega t}$. In many situations this is sufficient, but in the present case we need to be a little bit more careful. The starting point would be the text-book Fourier transform (or, perhaps rather a Laplace transform, as we are dealing with an initial-value problem, noting that this would also involve changing the lower limit in the integral below). Formally, we need 
\beq
\hat \chi (\omega, r)  = \int_{-\infty}^\infty \chi(t,r) e^{i\omega t} dt =- GM'\sum_{l\ge2} \sum_{m=-l}^l  W_{lm} r^l  Y_{lm} \left[  \int_{-\infty}^\infty {e^{i\omega t-im\psi(t)}  \over D^{l+1}(t)} dt\right]
\label{hatchi}\eeq
using hats to indicate frequency domain quantities.
In general, we have
\beq
\hat \chi (\omega, r)  = \sum_{l\ge2} \sum_{m=-l}^l  v_l r^l  f_{lm} (\omega) Y_{lm} 
\label{fdef}
\eeq
and in order to proceed we need to account for the evolution of the orbital separation $D$ and the phase $\psi$. The standard approach is to assume that the evolution is adiabatic and make use of the stationary phase approximation. In the simplest case one would progress by connecting a sequence of circular orbits with fixed separation \cite{l94}. Then we have
\beq
\dot \psi = \left[ {GM \over D^3}\right]^{1/2} = \Omega \quad  \longrightarrow \quad \psi \approx \Omega t
\eeq
and it follows straight away that
\beq
\hat \chi (\omega, r)  = \sum_{l\ge2} \sum_{m=-l}^l  v_l r^l  \delta(\omega - m\Omega) Y_{lm} 
\label{delta}
\eeq
As we are dealing with a linear problem, the delta function in frequency is inherited by all perturbed quantities and hence the solution will (at the end of the day) only have support at distinct frequencies $\omega = m\Omega$. In particular, there will be a distinction between the $m=0$ contribution, which is time independent and leads to a static tide, and  the dynamical contributions from the $m\neq0$ terms (usually discussed in terms of resonances associated with the star's oscillation modes \cite{l94,ks,wynn}).

The problem becomes more complicated when we account for the orbital evolution. The frequency support in \eqref{fdef} becomes less obvious and the inversion to the time domain more involved. However, this is the situation we are dealing with when we consider neutron star binaries sweeping through the sensitivity band of a gravitational-wave interferometer. To leading order, the evolution timescale due to gravitational radiation reaction is given by  
\beq
\dot D = - {64G^3 \over 5 c^5} {M^{4/3}\mathcal M^{5/3} \over D^3} 
\label{Deqn}
\eeq
where the dot is a time derivative and we have introduced the chirp mass
\beq
\mathcal M = \mu^{3/5} M^{2/5}
\eeq
with  the total mass $M = M_\star+M'$ and the reduced mass $\mu= M_\star M'/M$, as usual. The adiabatic approximation should be reasonable as long as
\beq
t_D = {D \over |\dot D|} \ll {1\over \Omega}
\eeq
and as long as we work at this level we can make progress in accounting for the shrinkage of the orbit (see, for example, the discussion in \cite{l94,ks}). However, in order to reach the precision required for gravitational-wave searches (and parameter extraction) we need to do better. The problem is messy as the orbital evolution requires  a higher order post-Newtonian analysis. Moreover, this may still not be sufficient, as the post-Newtonian scheme breaks down as the system approaches merger, leading to nonlinear simulations becoming necessary. Attempts to  match  approximate solutions and simulation results set the  (still developing) state of the art \cite{bern2,diet2,kawa,diet3,diet4,fouc}. However, the main point we need to appreciate is simple: A realistic model will not lead to \eqref{delta}. Rather, we end up with \eqref{fdef} which means that we have to account for a frequency dependent factor ($f$) in order to complete a dynamical description of the tidal response. This is a key part of the problem, but we will not attempt to solve it here. Instead, we take the view that the relevant frequency dependence (i.e. the function $f(\omega$)) is ``known" and focus on the response of the stellar fluid. Admittedly, this leaves the model incomplete but we nevertheless believe that our analysis adds valuable insight. 

\section{A simple homogeneous model}

Let us turn to the issue of solving \eqref{eul2} in the frequency domain. As a 
first example, we consider a  homogeneous, incompressible star. The calculation is straightforward and turns out to be instructive. We have $\Delta \hat \rho = 0$  and $\nabla_i \rho=0$, which means that the continuity equation reduces to 
\beq
\nabla_i \hat \xi^i  = 0 
\eeq
Moreover
\beq
\delta \rho  = \Delta \rho - \xi^j \nabla_i \rho  \longrightarrow \delta \hat \rho = 0 
\eeq

The Euler equation then simplifies to
\beq
-\omega^2 \hat  \xi_i +{1\over \rho}  \nabla_i \delta \hat p + \nabla_i \delta \hat \Phi = - \nabla_i \hat \chi
\label{eulhat0}
\eeq
while the Poisson equation becomes
\beq
\nabla^2 \delta \hat \Phi = 0 
\eeq
Since we also know that $\nabla^2 \hat \chi =0$, we see from \eqref{eulhat0} that
\beq
\nabla^2 \delta \hat p = 0 
\eeq

Expanding in spherical harmonics (here, and in the following, leaving out the implied, frequency dependent, factor of $f$ from \eqref{fdef} in all perturbation expressions), it is easy to see that the radial problem is degenerate (explicitly depending only on $l$, with the source term determining the $m$ dependence). Introducing notation such that (omitting the $m$ label on the different quantities for clarity)
\beq
\delta \hat \Phi = \sum_l \Phi_l Y_{lm}
\eeq
and similar for other perturbed quantities, 
the solution to the radial part of Laplace's equation for a given value of $l$ can be written (suppressing the indices on the coefficients, which should not cause confusion)
\beq
\Phi_l  = c r^l +{d \over r^{l+1}}
\eeq
That is, the (regular) interior solution is
\beq
 \Phi_l = c r^l
\eeq
Similarly, we have
\beq
  p_l = b r^l
\eeq
while \eqref{fdef} leads to
\beq
 \chi_l = v r^l
\eeq

The radial component of \eqref{eulhat0} leads to (after using the orthogonality of the spherical harmonics)
\beq
-\omega^2   \xi_l +{1\over \rho}\partial_r p_l + \partial_r    \Phi_l = - \partial_r  \chi_l
\label{eulhat2}
\eeq
(here and in the following steps we take $\xi_l$ to represent the radial component of the displacement)
or 
\beq
\omega^2   \xi_l = \left({1\over \rho} b + c + v \right) l r^{l-1} 
\label{xibal}
\eeq
We also know that the Lagrangian variation of the pressure must vanish at the surface, so
\beq
\Delta  p = \delta  p +  \xi^j \nabla_j p = 0 \quad \longrightarrow  \quad b r^l +  \xi_l p' = 0 \quad \mbox{at}\ r=R 
\eeq
Using hydrostatic equilibrium 
\beq
p' = -\rho \Phi' = - {GM_\star\rho \over R^3} r
\eeq
(with primes representing derivatives with respect to $r$) we have 
\beq
 b R^l - \xi_l {GM_\star\rho \over R^2} = 0 \longrightarrow {b R^{l-1}\over \rho} =   \xi_l {GM_\star \over R^3} 
\eeq

We also require continuity of the perturbed gravitational potential across the surface. This leads to
\beq
\Phi_l^\mathrm{in} (R) = \Phi_l^\mathrm{out} (R) = {d \over R^{l+1}}
\eeq
Meanwhile, the derivative of the potential must satisfy
(remembering that the density is not continuous at the surface in the case under consideration) 
\beq
(\Phi_l^\mathrm{in})' (R) = - (l+1) {d \over R^{l+2}}  -4\pi G\rho  \xi_l (R)
\eeq
Combining the two conditions, we see that the interior potential should satisfy
\beq
\Phi'_l + {l+1 \over r}  \Phi_l = -4\pi G\rho  \xi_l \quad \mbox{at} \quad r=R
\label{potcon1}
\eeq
That is, we have
\beq
(2l+1) c R^{l-1}= -4\pi G\rho \xi_l  \longrightarrow c R^{l-1} =   -{4\pi G\rho \over 2l+1}  \xi_l
\eeq 

In the absence of a tidal interaction, the problem reduces to that of free oscillations of the body. That is, the solution should provide the normal modes of the star. 
Thus, setting $v=0$ we can sanity check the calculation against the standard f-mode result for uniform density stars (see, for example, \cite{nagw}). From \eqref{xibal} we get
\beq
\omega^2  = l \left[   {GM_\star \over R^3}   -{3GM_\star \over (2l+1)R^3} \right] = {2l (l-1) \over 2l+1} {GM_\star \over R^3} \equiv \omega_f^2
\eeq

What changes when we add the tidal potential? In essence, we need  $\delta \hat \Phi \to \delta \hat \Phi + \hat \chi$ as  the total potential and its derivative have to be continuous at the surface. However,  the tidal potential is already continuous, so the condition \eqref{potcon1}  remains unchanged. This means that we have the two relations
\beq
(2l+1) c R^{l-1}= -4\pi G\rho   \xi_l  \longrightarrow   \xi_l= - {2l+1 \over 4 \pi G \rho} c R^{l-1}
\eeq 
(from before) and
\beq
\left( \omega^2   \xi_l  - {1\over \rho} b  l R^{l-1} \right) = \left( \omega^2  - l {GM_\star\over R^3} \right)  \xi_l   = \left( c + v \right) l R^{l-1} 
\label{xibal2}
\eeq
Combining these, we arrive at
\beq
 - (2l+1) c  \left( \omega^2  - l {GM_\star\over R^3} \right) =  {3GM_\star\over R^3} \left( c + v \right) l 
\eeq
Introducing the dimensionless frequency
\beq
\tilde \omega^2 = {\omega^2 \over GM_\star/R^3}
\eeq
and making use of the f-mode result, we arrive at the final relation
\beq
c = -{3l\over 2l+1} {v \over \tilde \omega^2 - \tilde \omega_f^2} = {\tilde \omega_f^2 - l \over \tilde \omega^2 - \tilde \omega_f^2} v
\label{cfinal}
\eeq

Let us now make contact with the Love number and the tidal deformability. In general, the matching of the gravitational potential at the star's surface provides the multipole moment, $I_l$, of the body according to 
\beq
 \Phi_l = - {4\pi G \over 2l+1} {I_{l} \over r^{l+1}} \quad \longrightarrow \quad d = c R^{2l+1} =  - {4\pi G \over 2l+1} I_l
\label{multip}
\eeq
If we also use
\beq
\chi_l = { 4\pi \over 2l+1} e_l r^l = v_l r^l
\eeq
then the Love numbers $k_l$ are defined as
\beq
G I_l = - 2k_l R^{2l+1} e_l \quad \longrightarrow \quad k_l =  {1\over 2} { \Phi_l(R) \over \chi_l(R)} =  {c \over 2v}
\label{love1}
\eeq
For the homogeneous model we then have (cf. the discussion in  \cite{ogilvie})
\beq
k_l =  {\tilde \omega_f^2 - l \over  2(\tilde \omega^2 - \tilde \omega_f^2)} 
\label{kleff1}
\eeq

The model is admittedly simplistic, but it has led us to a useful result. The final relation \eqref{kleff1} links the tidal deformability, expressed in terms of the Love number, to the frequency of the fundamental mode of the star. We can now make a number of observations:
First of all, in the limit $\tilde \omega\ll \tilde \omega_f$ we have
\beq
k^\mathrm{eq}_l = {1\over 2} {3l\over 2l+1} {1\over \omega_f^2}=  {1\over 2} {3l\over 2l+1} {2l+1\over 2l (l-1)} = {3\over 4 (l-1)}
\eeq
This is the expected result for the ``equilibrium'' tide, as we arrive at the same answer by setting $\omega=0$ from the outset \cite{willpo}.
Secondly, we see how the Love number depends on the f-mode frequency. This helps explain phenomenological relations like those developed in \cite{chan}. Next, we note that, in the case of a fixed orbital distance, the time-domain result splits into the expected static contribution (from $m=0$) and a  time dependent component with the f-mode resonances at $\omega_f = \pm m \Omega$ (from the $m\neq 0$ terms) \cite{holai,l94,ks,wynn}. However, the difference is entirely due to the frequency dependence of $f(\omega)$ which comes into play when we invert the transform to the time domain. Finally, and perhaps most notably, the relation \eqref{kleff1}  provides the frequency dependence of the star's tidal response. This is a key ingredient for a model involving evolving orbits and, as we will see,  crucial for an understanding of the role the matter composition plays in the problem.

\section{The effective Love number}

Inspired by the homogeneous model, let us now consider the  problem for more realistic stellar models. In essence, we need to solve
\beq
-\omega^2  \hat \xi_i +{1\over \rho}  \nabla_i \delta \hat p - {1\over \rho^2} \delta \hat \rho \nabla_i p + \nabla_i \delta \hat \Phi = - \nabla_i \hat \chi
\label{eul2b}
\eeq
for some given equation of state. 

First of all, recalling the usual argument for the equilibrium tide, 
 we assume that the matter is in chemical equilibrium (we will relax this assumption later). That is, we take the equation of state is to be barotropic,  $p=p(\rho)$, such that 
\beq
\delta p=  \left({\partial p\over \partial \rho}\right)_\mathrm{\beta} \delta \rho = c_s^2 \delta \rho  
\label{sound}
\eeq
with $c_s^2$  the speed of sound (calculated for matter in beta equilibrium).
We also know that the unperturbed background configuration is such that 
\beq
\nabla_i p = - \rho \nabla_i \Phi = - \rho g \quad \longrightarrow  \quad p' = - \rho \Phi' 
\label{baro}
\eeq
where (as before) a prime indicates a radial derivative  and we have introduced the gravitational acceleration $g$.

Expanding in spherical harmonics (as before) and introducing
\beq
U_l =  \Phi_l + \chi_l
\eeq
we have the radial component of the Euler equation (in the $\omega\to 0$ limit)
\beq
 p'_l - {p'\over\rho} \rho_l = - \rho U'_l
 \label{rad}
\eeq
and the angular part 
\beq
p_l = - \rho U_l
\label{ang}
\eeq
Meanwhile,  the perturbed  Poisson equation becomes
\beq
r^2U^{''}_l + 2r U'_l - l(l+1)U_l = 4\pi G r^2 \rho_l
\label{pois}
\eeq
where we have made use of the fact that $\chi_l$ solves the corresponding homogeneous equation.

Let us now pause to note that the problem appears to be over determined. We seem to have too many equations for the number of variables. 
However, taking a radial derivative of \eqref{ang} we get
\beq
p'_l = - \rho' U_l - \rho U'_l  = {\rho'\over \rho} p_l - \rho U'_l 
\eeq
Using this in \eqref{rad} we have
\beq
{\rho'\over \rho} p_l - {p'\over\rho} \rho_l = 0 \longrightarrow \rho' p_l = p' \rho_l 
\label{barot}
\eeq
which is  consistent with \eqref{sound} as long as we have a barotropic equation of state. This identity reduces the number of equations, so the problem is  well posed, after all. In essence, this is the Newtonian version of the result discussed in \cite{penner}. 

Now we have 
\beq
r^2 U^{''}_l + 2r U'_l + \left[{4\pi G r^2 \rho \over c_s^2} - l(l+1) \right] U_l = 0
\label{maineq}
\eeq
which is easily solved by integrating from the centre to the surface of the star. At the surface we match to the exterior potential. This matching provides the multipole moment, $I_l$, and the Love number, $k_l$, reproducing the steps from \eqref{multip} to \eqref{love1}. 

However, as we have already indicated, the  calculation of the tidal deformability and the Love number may not be (strictly) valid for realistic neutron star binaries close to merger. Basically, the nuclear reactions required to establish chemical equilibrium are too slow to act on the inspiral timescale. If this is the case we can no longer assume that the equation of state for the perturbations is barotropic. Instead, it would be  reasonable to assume that the composition is held frozen as the system sweeps through the sensitivity band of a ground based detector. This changes the response of the stellar fluid to the tidal driving which, in turn, allows us to estimate the impact   the matter composition has  on the problem. 

\begin{figure}
\begin{center}
\includegraphics[width=0.6\textwidth]{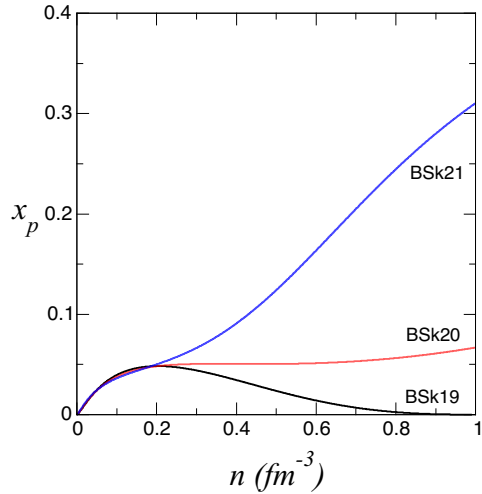}
\end{center}
\caption{Proton fractions for three ``realistic'' BSk models from \cite{Bsk1,Bsk2}. } 
\label{fractions}
\end{figure}

Taking the matter composition into account (assuming a simple model composed of neutrons, protons and electrons),
we  have a two-parameter equation of state $p=p(\rho,x_\p)$ (say), where $x_\p=n_\p/n$ is the proton fraction. In the Newtonian context the mass density is simply $\rho = m_B n$ where $n$ is the baryon number density. Hence, we can think of $\rho$ as a proxy for the number density. Moreover,  the continuity equation \eqref{continuity} remains unchanged (although, strictly speaking, it now represents baryon number conservation). As an illustration of the range of values we may need to consider, and the corresponding variation with density, we illustrate the proton fraction for three equation of state parametrisations from the Brussel-Montreal collaboration \cite{Bsk1,Bsk2} in figure~\ref{fractions}. The BSk models have the particular advantage that one can readily work out the different thermodynamical derivatives we require.

In order to account for nuclear reactions, we follow \cite{gmode} and introduce a new dependent variable $\beta=\mu_\n-\mu_\p-\mu_\e$ which encodes the deviation from chemical equilibrium (with $\mu_\x$, $\x=\n,\p,\e$ being the chemical potentials). We then have
\beq
\Delta p = \left( {\partial p \over \partial \rho} \right)_{\beta} \Delta \rho +  \left( {\partial p \over \partial \beta} \right)_{\rho} \Delta \beta 
= c_s^2  \Delta \rho +  \left( {\partial p \over \partial \beta} \right)_{\rho} \Delta \beta 
\label{eqone}
\eeq
with 
\beq
\Delta \beta = {\mathcal B \over 1 + i \mathcal A/\omega} \Delta \rho
\label{dbeta}
\eeq
and the thermodynamical derivatives
\beq
\mathcal A =  \left( {\partial \beta \over \partial x_\p}\right)_{\rho} {\gamma \over n} \ , \qquad 
\mathcal B = \left( {\partial \beta \over \partial \rho}\right)_{x_\p}
\eeq
The relevant reaction rate is encoded in $\gamma$, assuming a small deviation from equilibrium.

Let us consider the relevant timescales. Introducing a characteristic reaction time 
\beq
t_R = {1 \over |\mathcal A|}  
\eeq
we see that, if the reactions are fast compared to the dynamics (assumed to take place on a timescale $\sim 1/\omega$) then we have $\omega t_R  \ll 1$ and if we  consider the evolution in the adiabatic limit (where $\omega\to 0$) then 
\beq
\Delta \beta \approx 0 \ .
\eeq
The system remains in beta-equilibrium and the standard (barotropic) analysis of the tidal deformability holds. 

However, we are interested in the opposite limit -- where reactions are slow.  Then we have $\omega t_R  \gg 1$ (see \cite{gmode}) and we can Taylor expand \eqref{dbeta} to get
\beq
\Delta \beta \approx \mathcal B \left(  1-  i \mathcal A/\omega \right)  \Delta \rho \approx \mathcal B \Delta \rho
+ \mathcal O\left( {1\over \omega t_R}\right)
\label{dbeta}
\eeq
However, we can no longer  (at least not meaningfully) take the $\omega \to 0$ limit. If we insist on doing this,  the problem  inevitably becomes barotropic as, for any fixed $t_R$, we must cross over into the fast reaction regime. The argument is analogous to that for the g-modes, discussed in \cite{gmode}. In essence, if we want to consider the impact of a frozen composition we cannot work in the static limit -- we have to consider dynamical aspects of the tide and  solve the frequency dependent problem. 

\begin{figure}
\begin{center}
\includegraphics[width=\textwidth,clip]{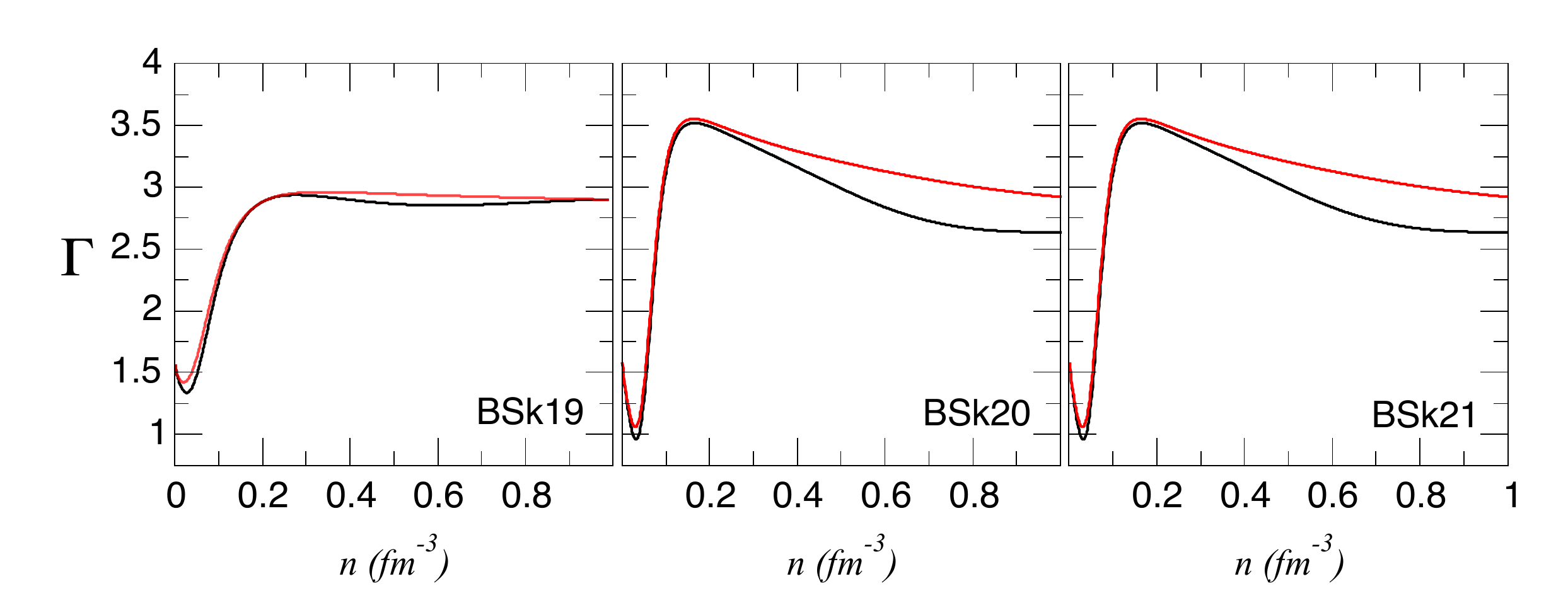}
\end{center}
\caption{Adiabatic indices for the three BSk models from figure~\ref{fractions} (black = $\Gamma$ and red=$\Gamma_1$). } 
\label{gammas}
\end{figure}

\begin{figure}
\begin{center}
\includegraphics[width=10cm,clip]{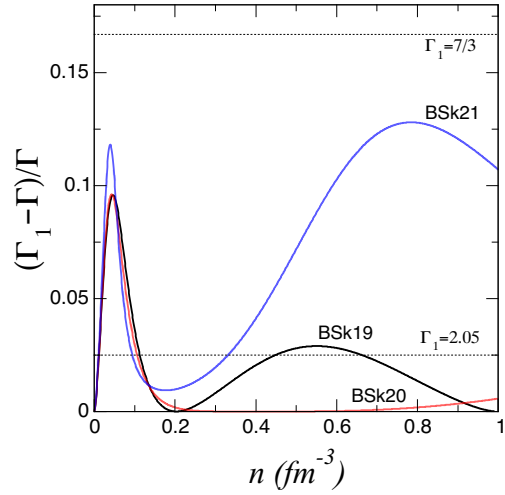}
\end{center}
\caption{Difference between frozen composition and barotropic adiabatic indices for the models from figure~\ref{gammas}. } 
\label{diffs}
\end{figure}

Motivated by this conclusion, let us see how the analysis changes if we consider a general (compressible and possibly stratified) model. First of all, we need to quantify the difference between the barotropic and the frozen-composition cases. Effectively, this can be done by noting that we have
\beq
\Delta p = \Gamma_1 \Delta \rho
\eeq
with 
\beq
\Gamma_1 = \Gamma \equiv \left( {\partial p \over \partial \rho} \right)_\beta
\eeq
in the barotropic case, but with $\Gamma_1$ distinct from $\Gamma$, in general. An illustration, again for three BSk models, of the difference between the two adiabatic indices (in the two limits we have discussed) is provided in figure~\ref{gammas}. Meanwhile, the relative difference between $\Gamma_1$ and $\Gamma$ is shown in figure~\ref{diffs}. In the numerical examples considered later, we will take $\Gamma_1$ to be constant, with values spanning the range suggested by figure~\ref{gammas}. This is, perhaps, not particularly realistic, but it serves as a useful starting point.

\subsection{Mode expansion}

Taking a lead from the results for the homogeneous model, we aim to express  the driven response of the fluid in terms of a set of normal modes\footnote{Note that, while it is common to assume that the modes form a complete set,  it is not clear that this is actually the case for more realistic neutron star models.}, corresponding to solutions $\xi_n$ (where $n$ is a label that identifies the modes, say in terms of the number of nodes in the radial eigenfunction and the corresponding spherical harmonics). Letting the (real) mode-frequency be $\omega_n$ we have (leaving out the hats for frequency domain quantities in order to avoid the notation becoming cluttered)
\beq
 \xi^i = \sum_n a_n   \xi_n^i 
\eeq
and each individual mode (labelled by $n$) satisfies
\beq
-\omega_n^2 \rho    \xi_n^i  + C  \xi_n^i=0
\eeq
where the $C$ operator is messy but we do not need an explicit expression here.

Making use of the inner product from \cite{fs78b}
\beq
\langle  \xi_{n'} , \rho  \xi_n \rangle = \int \rho  \xi^*_{n'} \xi_n d^3 x 
\eeq
(where the asterisk indicates the complex conjugate)
it is easy to show that the modes are orthogonal (at least as long as  the frequencies are real). That is, we have (keeping the normalisation of the modes explicit for the moment)
\beq
\langle \xi_{n'} , \rho \xi_n \rangle = A_n^2 \delta_{n n'}
\eeq
We  use the orthogonality to rewrite \eqref{eul2b} as an  equation for the mode amplitudes:
\beq
\ddot a_n + \omega_n^2 a_n = - {1\over A_n^2} \langle \xi_n, \rho \nabla \chi\rangle
\eeq
Finally, making use of the perturbed continuity equation
\beq
\delta \rho_n = - \nabla_i (\rho \xi_n^i)  
\eeq
and integrating by parts, we have (assuming that the density vanishes at the surface of the star)
\beq
- \langle \xi_n, \rho \nabla\chi\rangle = - \int \rho (\xi^i_n)^* \nabla_i \chi d^3x =  \int \chi  \nabla_i (\rho \xi^i_n)^* d^3 x = - \int \chi \delta \rho^*_n d^3x 
\eeq

In general, e.g. when the star is spinning, it may be practical to express the stellar perturbations with respect to a different set of spherical harmonics \cite{holai}, but we will not worry about this here. 

Making use of the given expression for the tidal potential \eqref{fdef}
we have an equation for the driven modes (for each $l$, as the different values of $m$ are still degenerate)
\beq
-\omega^2 a_n + \omega_n^2 a_n =   v_{l}   Q_n
\eeq
where we have introduced the ``overlap integral'' 
\beq
Q_{n} = - \int \delta \rho^*_n r^{l+2}  dr
\label{overlap}
\eeq
In effect, we have a driven set of modes
with amplitude
\beq
a_n =   {1 \over \omega_n^2 - \omega^2}  {Q_{n} \over A_n^2}  v_{l}
\label{reason}
\eeq
which may become resonant during a binary inspiral. This is a well-known result \cite{l94,ks,pt77}.

Turning to the matching at the surface, the perturbed gravitational potential satisfies
\beq
r^2 {d^2 \over dr^2} \delta  \Phi  + 2r {d \over dr} \delta  \Phi - l(l+1)  \delta  \Phi = 4\pi G r^2 \delta  \rho 
\label{pois}
\eeq

For the overlap integral we need
\begin{multline}
r^{l+2} \delta  \rho= {r^l\over 4\pi G}  \left[ r^2 {d^2 \over dr^2} \delta  \Phi  + 2r {d \over dr} \delta  \Phi - l(l+1)  \delta  \Phi \right] \\
= {r^l\over 4\pi G}  \left[ {d\over dr} \left( r^2 {d \over dr} \delta  \Phi \right) - l(l+1)  \delta  \Phi \right]
\end{multline}
Integrating by parts, we get
\begin{multline}
4\pi G \int_0^R r^{l+2} \delta  \rho dr = \left[ r^l  \left( r^2 {d \over dr} \delta  \Phi \right) \right]_0^{r=R} - \int_0^R l r^{l+1} \left(  {d \over dr} \delta  \Phi \right) dr  \\
-l(l+1) \left[ {1\over l+1} r^{l+1}  \delta  \Phi \right]_0^R +  l \int_0^R r^{l+1} \left( {d\over dr}  \delta  \Phi \right) dr \\
=  R^{l+2} \left[{d \over dr} \delta  \Phi \right]_{r=R} 
-l  R^{l+1}  \delta  \Phi (R) 
\label{intpart}
\end{multline}

However, we know that the solution should satisfy (since the density now vanishes as $r\to R$) 
\beq
{d \over dr} \delta  \Phi + {l+1 \over r} \delta  \Phi =0 \quad \mbox{at} \quad r=R
\label{potcon}
\eeq
so \eqref{intpart} leads to
\beq
4\pi G \int_0^R r^{l+2} \delta  \rho dr = - (2l+1) R^{l+1}  \delta  \Phi (R) 
\eeq
and we have 
\beq
Q_{n}  =   { 2l+1 \over 4\pi G } R^{l+1}  \delta  \Phi_n (R) =  I_n
\eeq
We recognize $I_n$ as the contribution each mode makes to the mass multipole moment.

\subsection{The effective tidal deformability}

We now want  to connect the mode expansion to the tidal deformability and the effective Love number. The purpose is to discuss what happens far away (above and below) from a given resonance. In order to do this, we need a representation of the perturbed gravitational potential associated with the mode expansion.

In order to obtain the desired result, we need to connect the mode expansion for the displacement to the gravitational potential. It is then useful to consider the components of the displacement vector 
\beq
\xi^i =  {W(r) \over r }\nabla^i r  + V(r) \nabla^i Y_{lm} 
\eeq
and similarly for the contribution from each mode
\beq
\xi_n^i = \left[ \left(W_n {\nabla^i r  \over r }\right) Y_{lm} + V_n \nabla^i Y_{lm}  \right]
\eeq
from which it follows that the normalisation constant is given by
\beq
A_n^2 = \int_0^R \rho \left[ W_n^2 + l(l+1) V_n^2\right] dr
\eeq
We now see that the tidal problem leads to a fluid displacement of form
\beq
\xi^i =  \sum_n {1 \over \omega_n^2 - \omega^2} {Q_n \over A_n^2} \left[  \left(W_n {\nabla^i r  \over r }\right) Y_{lm}   + V_n \nabla^i Y_{lm}  \right]
\eeq

Going back to  the Euler equation, the $\theta$-component leads to
\beq
-\omega^2  V + {1\over \rho} \delta p  + \delta \Phi = - \chi\  (= - v_{l} r^l)
\eeq 
Moreover, at the surface we have $\Delta p = 0$ so
\beq
-\omega^2  V(R) -{ p' \over \rho}  {W(R) \over R}  + \delta \Phi(R) = -\chi(R)
\eeq 
and it follows that
\beq
\delta \Phi(R) = - \chi(R) + \omega^2 V(R) - g  {W(R) \over R}
\eeq
That is, we have 
\beq
k_l = {1\over 2} {\delta \Phi(R) \over \chi(R) } = -{1\over 2} + {1\over 2 v_l R^l} \left[ \omega^2 V(R) - g {W(R)\over R} \right]
\eeq
Making use of the mode expansion, this becomes
\beq
k_l =  -{1\over 2} + {1\over 2 R^l} \sum_n {Q_n \over A_n^2} {1 \over \omega_n^2 - \omega^2}\left[ \omega^2 V_n(R) - {GM\over R^3} W_n(R)\right]
\label{keff1}
\eeq

In the low-frequency limit (with $\omega$ much smaller than the lowest frequency mode\footnote{This is a somewhat subtle issue, but the discussion of the g-mode spectrum in \cite{gmode} ensures that the limiting procedure makes sense.}), we have
\beq
k_l \approx -{1\over 2} - {1\over 2 R^l} \sum_n {Q_n \over  \omega_n^2 A_n^2 } {GM_\star\over R^3} W_n(R) \\
= -{1\over 2} - {1\over 2} \sum_n {Q_n \over  \tilde \omega_n^2 A_n^2} { W_n(R) \over R^l}
\eeq
where we have used the dimensionless frequency from before.

Now recall that
\beq
Q_{n}  =   { 2l+1 \over 4\pi G } R^{l+1}  \delta  \Phi_n (R)
\eeq
and use 
\beq
\delta \Phi_n(R) =   \omega_n^2 V_n(R) - g  {W_n(R) \over R} 
\eeq
to get 
\beq
Q_{n}  =  { 2l+1 \over 4\pi G } R^{l+1}    \left[  \omega_n^2 V_n(R) - g  {W_n(R) \over R} \right] 
\label{Qrel}
\eeq
This  leads to the final result
\beq
k_l = -{1\over 2} - {2l+1\over 8\pi  }  \sum_n {M_\star  A_n^2\over  \tilde \omega_n^2} { W_n(R) \over R^2}  \left[  \tilde \omega_n^2 V_n(R) -   W_n(R)\right] 
\label{kfinal}
\eeq

Alternatively, we can rewrite \eqref{Qrel} as
\beq
W_n(R) = - {4\pi\over 2l+1} {Q_n \over M_\star R^{l-2}} \left[ 1 - \tilde \omega_n^2 \left( {V_n\over W_n}\right)_R\right]
\eeq
Introducing the dimensionless overlap integral \cite{ks}
\beq
\tilde Q_n = {Q_n \over M R^l}
\eeq
we have
\beq
{W_n(R) \over R^l}  = - {4\pi\over 2l+1} {\tilde Q_n \over R^{l-2}} \left[ 1 - \tilde \omega_n^2 \left( {V_n\over W_n}\right)_R\right]^{-1}
\label{WnR}
\eeq
and
\beq
k_l \approx -{1\over 2} + {2\pi \over 2l+1} \sum_n {\tilde Q_n^2 \over  \tilde \omega_n^2 } \left( { M_\star R^2 \over A_n^2} \right) \left[ 1 - \tilde \omega_n^2 \left( {V_n\over W_n}\right)_R\right]^{-1}
\eeq
and if we (finally) normalise the modes in such a way that 
\beq
A_n^2 = M_\star R^2
\eeq
we arrive at the expression
\beq
k_l \approx -{1\over 2} + {2\pi \over 2l+1} \sum_n {\tilde Q_n^2 \over  \tilde \omega_n^2 }  \left[ 1 - \tilde \omega_n^2 \left( {V_n\over W_n}\right)_R\right]^{-1}
= - {1\over 2} + \sum_n k_l^n 
\label{klfinal}
\eeq

\begin{figure}
\begin{center}
\includegraphics[width=0.5\textwidth]{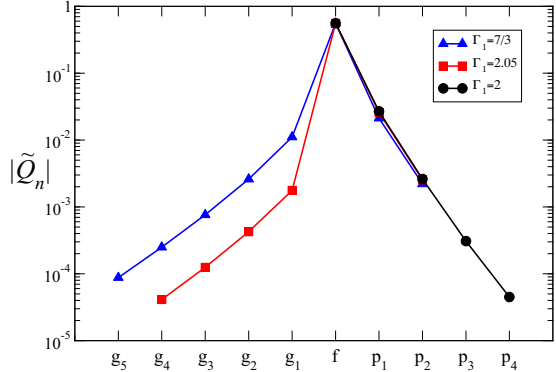}
\end{center}
\caption{A summary of the results for the overlap integrals $|\tilde Q_n|$ for the three models considered in tables~\ref{table1}-\ref{table3}. The results show that, while the matter composition has little impact on the fundamental f-mode and the pressure p-modes, the impact of the gravity g-modes is more pronounced for models with $\Gamma_1$ deviating significantly from the background adiabatic index $\Gamma$.} 
\label{overlaps}
\end{figure}

As a quantitative test of \eqref{klfinal} we compare results for three models corresponding to a background configuration with $\Gamma=2$ (i.e. a standard $n=1$ polytrope). Our reference model is barotropic, $\Gamma_1=\Gamma$, and we compare it to two stratified models, with $\Gamma_1=2.05$ and $7/3$, respectively. The results from figure~\ref{diffs} suggest that these cases span the range of ``reasonable'' parameter values. Moreover, mode results and overlap integrals for the same values of $\Gamma_1$ are already available from \cite{ks}, which means that we have an independent test of the numerics. However, we need to go one step further and add the ratio of the two eigenfunctions at the surface. The numerical results, listed in tables~\ref{table1}-\ref{table3} (see also figure~\ref{overlaps}), demonstrate the relative importance of the g-modes for strongly stratified models, evident from the larger values of the overlap integrals $\tilde Q_n$ and the relative contributions ($k_l^n$) to the overall Love number.  

\begin{table}
\begin{tabular}{l c c c c}\hline
mode & $\tilde \omega_n$ & $|\tilde Q_n|$ & $(V_n/W_n)_R$ & $k_l^n$ \\ 
\hline
$p_4$ 	& 9.0471  		& $4.4927\times 10^{-5}$ 	& $1.2218\times10^{-2}$	& $-1.7426\times 10^{-6}$\\ 
$p_3$ 	& 7.2581  		& $3.0791\times 10^{-4}$ 	& $1.8983\times10^{-2}$	& $-5.9238\times 10^{-4}$\\ 
$p_2$ 	& 5.4144 		& $2.6168\times 10^{-3}$ 	& $3.4104\times10^{-2}$	& $1.3425\times 10^{-3}$\\ 
$p_1$ 	& 3.4615  		& $2.6888\times 10^{-2}$ 	&  $8.3845\times10^{-2}$	& $-1.6411\times 10^{-2}$ \\ 
$f$ 		& 1.2267 		& $5.5792\times10^{-1}$ 	& $4.4163\times10^{-1}$	& 0.77528\\ 
\hline
\end{tabular} 

\caption{Numerical results for barotropic $n=1$ polytropes (with $\Gamma_1=2$) and $l=2$.}
\label{table1}
\end{table}

\begin{table}
\begin{tabular}{l c c c c}\hline
mode & $\tilde \omega_n$ & $|\tilde Q_n|$ & $(V_n/W_n)_R$ & $k_l^n$ \\ 
\hline
$p_2$ 	& 5.4949  	& $2.5459\times 10^{-3}$ 	& $3.3111\times10^{-2}$	& $1.6522\times 10^{-3}$\\ 
$p_1$ 	& 3.5204  	& $2.5865\times 10^{-2}$ 	&  $8.1049\times10^{-2}$	& $-1.9849\times 10^{-2}$ \\ 
$f$ 		& 1.2274  		& $5.5796\times10^{-1}$ 	& $4.4009\times10^{-1}$	& 0.77055\\ 
$g_1$ 	& 0.1845  		& $1.7657\times 10^{-3}$ 	& 27.8841	& $2.2458\times 10^{-3}$ \\ 
$g_2$ 	& 0.1270  		& $4.264 \times 10^{-4}$ 	& 60.9918 & $8.2137\times 10^{-4}$ \\ 
$g_3$ 	& 0.0975  		& $1.2493\times 10^{-4}$ 	& 104.4819	& $3.3549\times10^{-4}$ \\ 
$g_4$ 	& 0.0794  		& $4.1194\times 10^{-5}$ 	& 158.1560	& $1.4542\times10^{-4}$ \\ 
\hline
\end{tabular} 

\caption{Numerical results for $n=1$ polytropes with $\Gamma_1=2.05$ and $l=2$. }
\label{table2}\end{table}

\begin{table}
\begin{tabular}{l c c c c}\hline
mode & $\tilde \omega_n^2$ & $|\tilde Q_n|$ & $(V_n/W_n)_R$ & $k_l^n$ \\ 
\hline
$p_2$ 	& 5.9316  	& $2.2079\times 10^{-3}$ 	& $2.8417\times10^{-2}$ & $9.6779\times 10^{-4}$\\ 
$p_1$ 	& 3.8411  	& $2.1197\times 10^{-2}$ 	&  $6.8024\times10^{-2}$	& $-1.0631\times 10^{-2}$ \\ 
$f$ 		& 1.2291  		& $5.5805\times10^{-1}$ 			& $4.3227\times10^{-1}$	& 0.74685\\ 
$g_1$ 	& 0.4361  		& $1.110\times 10^{-2}$ 	& 4.9584 	& $1.4299\times 10^{-2}$ \\ 
$g_2$ 	& 0.3035  		& $2.6048 \times 10^{-3}$ 	& 10.6541	& $5.0402\times 10^{-3}$ \\ 
$g_3$ 	& 0.2343 		& $7.6158\times 10^{-4}$ 	& 18.0985 	& $2.0379\times10^{-3}$ \\ 
$g_4$ 	& 0.1912 		& $2.4981\times10^{-4}$ & 27.2719 & $8.7876\times10^{-4}$  \\
$g_5$ 	& 0.1618 		& $8.7649\times10^{-5}$ & 38.1746 & $3.9289\times10^{-4}$  \\
\hline
\end{tabular} 

\caption{Numerical results for $n=1$ polytropes with $\Gamma_1=7/3$ and $l=2$. }
\label{table3}
\end{table}

\begin{table}
\begin{tabular}{|c c | c c | c c|}
\multicolumn{2}{c}{$\Gamma_1=2$} & \multicolumn{2}{c}{$\Gamma_1=2.05$}  & \multicolumn{2}{c}{$\Gamma_1=7/3$} \\
mode &  $k_l$ & mode &  $k_l$ & mode &  $k_l$\\ 
\hline
$f$		& 0.27528	&	$f$ 		& 0.27055 & $f$ 			& 0.24685\\ 
$+p_1$	& 0.25887	&	$+p_1$ 	& 0.25526 & $+g_1$ 	& 0.26115 \\ 
$+p_2$ 	& 0.26021		&$+p_2$ 	& 0.25653 & $+p_1$ 	& 0.25052\\ 
$+p_3$	& 0.26015		&$+g_1$ 	& 0.25878 & $+g_2$ 	& 0.25556\\ 
		&		&$+g_2$ 	& 0.25960 & $+p_2$ 	& 0.25653\\  
		&		&$+g_3$ 	& 0.25993 & $+g_3$ 	& 0.25856 \\ 
		&		&$+g_4$ 	& 0.26008 & $+g_4$ 	& 0.25944 \\ 
		&		&   		& 		& $+g_5$ 		& 0.25983 \\ 
\hline
& 	$9\times10^{-4}$	& 	& $7\times10^{-4}$	& 	& $3\times10^{-4}$\\
\hline
\end{tabular} 

\caption{The accumulated contribution to the tidal deformability from the different modes, in order of relevance of the contribution, for the three different models we consider. We expect (for the barotropic $\Gamma=2$ case) to have $k_l\approx 0.259909$. If we add up the contributions from the different modes in each case, the results converges to the expected answer. The mode sum is  always dominated by the f-mode with a modest correction from the other modes of the star, but the enhanced importance of the g-modes with increasing stratification is notable.}
\label{table4}
\end{table}

The numerical results are interesting. In each case, the mode sum converges to the expected value for the Love number. For a barotropic model with $\Gamma=2$ the results should be $k_l\approx 0.259909$ \cite{willpo} and the results from table~\ref{table1} do, indeed, converge towards this number. This tells us that the sum over the star's different oscillation modes provides an alternative representation for the Love number. It is important to note that this is true also for the stratified models, see table~\ref{table4}. The fundamental f-mode provides the dominant contribution in all cases (as expected given that this mode most closely resembles the tidal driving force), but in order to have a precise representation we need to account for both pressure and gravity modes. In fact, for the (somewhat extreme) $\Gamma_1=7/3$ case the first g-mode is more important than the p-modes.

Should we be surprised about these results? Probably not. As long as the modes of the star are complete, the mode-sum representation of the Love number must converge to the barotropic answer (after all, we can determine $k_l$ in the static limit).  The contribution of individual modes changes with $\Gamma_1$, but the final answer is always the same. Nevertheless, the results are important. They provide the first actual demonstration that the dynamical response of the star, usually discussed in terms of mode resonances \cite{l94,ks,wynn}, has as its static limit the tidal deformability. In essence, we have an explicit link between the tidal deformability and asteroseismology. The results also point us in the direction we have to go if we want to quantify the impact of matter composition on the tidal response. We need to consider the dynamics encoded in the frequency dependence of \eqref{keff1}.

\subsection{The dynamical tide}

\begin{figure}
\begin{center}
\includegraphics[width=0.95\textwidth]{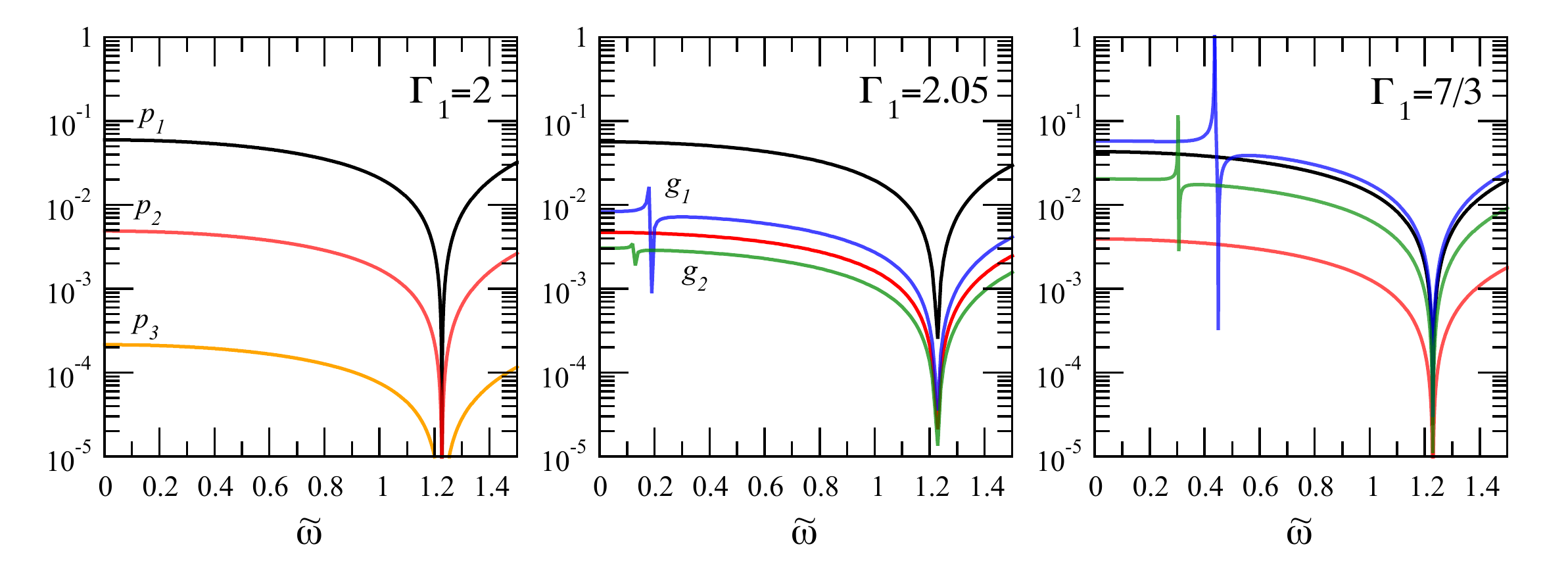}
\end{center}
\caption{Relative contributions to the tidal deformability (compared to that of the f-mode alone). The three panels show, from left to right: the barotropic case $\Gamma_1=\Gamma=2$, $\Gamma_1=2.05$ and $\Gamma_1=7/3$. Individual modes are colour coded (as indicated in the panels), with the same colour representing the same mode in all panels.} 
\label{relerr}
\end{figure}

Having demonstrated that the sum over the star's oscillation modes provides a precise description of the tidal response in the static limit, let us turn to the dynamical response associated with finite frequencies. The static result gives us confidence that this is a sensible exercise, obviously closely related to previous discussions of resonant mode excitation in binary systems \cite{l94,ks} but at the same time adding insight into the role of the (no longer static) $m=0$ contribution. Taking  \eqref{keff1} as our starting point and introducing dimensionless frequencies, as before, we have
\beq
k_l =  -{1\over 2} - {1\over 2 R^l} \sum_n {Q_n \over A_n^2}  { W_n(R) \over \tilde \omega_n^2 - \tilde \omega^2}\left[1- \tilde \omega^2 \left({V_n \over W_n }\right)_R\right]
\eeq
Next we use \eqref{WnR} and normalise the modes to arrive at 
\beq
k_l =  -{1\over 2} + {2\pi\over 2l+1} \sum_n {\tilde Q_n^2  \over \tilde \omega_n^2 - \tilde \omega^2}\left[1- \tilde \omega^2 \left({V_n \over W_n }\right)_R\right]   \left[ 1 - \tilde \omega_n^2 \left( {V_n\over W_n}\right)_R\right]^{-1}
\label{efftide}
\eeq
This is the final result. It provides a closed expression for the frequency dependent tidal response (encoded in $k_l$) and, given the numerical results from tables~\ref{table1}-\ref{table3} it is straightforward to obtain the desired dynamical behaviour. This, in turn, allows us to quantify the level at which each individual mode contributes to the overall result. Results for the three different values of $\Gamma_1$ we have considered are presented in figure~\ref{relerr}. The different panels show the relative contributions to the tidal deformability (compared to that of the f-mode alone). The resonances associated with each mode, which occur when $\tilde \omega = \tilde \omega_n$, are easily distinguishable in each case and the resonance associated with the f-mode leads to a common feature (at $\tilde \omega \approx 1.2$) in all panels. The results tell us that modes other than the f-mode contribute to the overall result at the few percent level. Moreover, the results for $\Gamma_1=7/3$ bring out the fact that, in this case the leading g-mode dominates over the first p-mode for all frequencies. However, for the (likely) more realistic case with $\Gamma_1=2.05$ the g-mode contribution is almost an order of magnitude weaker than that of the first p-mode. These results are important as they provide the first (we believe) demonstration of the level at which frozen matter composition impacts on the tidal response across the range of frequencies relevant for a binary inspiral. Of course, the notion that resonant mode-excitation comes into play is not at all new. However, the discussion tends to focus on $m\neq0$ modes (which would make a dynamical contribution for systems with a fixed orbital separation). The fact that the $m=0$ contribution has similar features (which should come into play for evolving orbits, when the function $f(\omega)$ is non-trivial, see the discussion in section~2) appears to not have been appreciated previously.

\begin{figure}
\begin{center}
\includegraphics[width=0.6\textwidth]{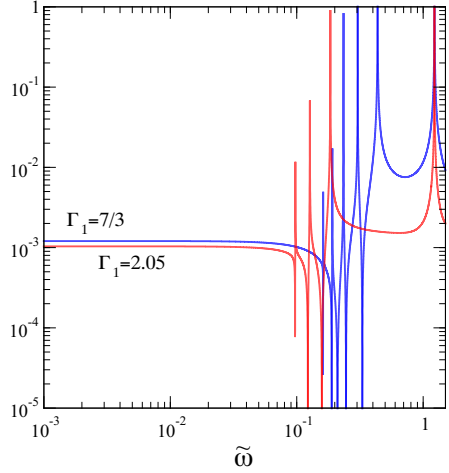}
\end{center}
\caption{Relative difference between the two stratified models and the barotropic tidal deformability as functions of the dimensionless frequency $\tilde\omega$.} 
\label{totalsum}
\end{figure}

Finally, the overall mode contribution to the tidal response (including only the modes listed in tables~\ref{table1}-\ref{table3}) is illustrated in figure~\ref{totalsum}. This figure requires some additional explanation. We know from table~\ref{table4} that the modes we include lead to a truncation error at the $10^{-3}$ level. This sets the level for the horizontal part of the curves towards the left of the figure. The inclusion of further modes should bring this level down (up to numerical precision). Instead, we should focus on the behaviour at frequencies beyond the first mode resonance. This  illustrates the difference between the barotropic models and the stratified cases, again suggesting that matter composition may affect the result at the percent level.

\section{Implications}

We have discussed the tidal response of a neutron star during the late stages of neutron star binary inspiral. In particular, we have focussed on the role of the matter composition. This issue has previously been ignored as studies have almost exclusively focussed of barotropic fluid models.  However, it is natural to argue (given the timescale involved) that the matter composition should remain ``frozen'' during the late stages of binary inspiral, leading to a stratified perturbation problem (where the adiabatic index of the perturbation is different from that of the equilibrium background). This connects with previous work on tidal resonances, which has quantified the role of the g-modes (which obviously rely on stratification for their existence). Similarly, the proposed nonlinear pg-instability \cite{arras,wein,essick,ligo_pg} relies on the coupling between p-modes and g-modes. Of course, one may argue that the impact of stratification on the tidal response should be small enough that it can safely be ignored. Indeed, our numerical results indicate that the difference is at (or below) the level of a few percent. However, it is nevertheless important to quantify this contribution. We need to do this in order to understand systematic ``errors'' associated with the assumed physics, which ultimately determines the accuracy with which we can hope to extract stellar parameters like the radius from observations. Today's gravitational-wave detectors are not at a level where a change of a percent in the tidal response makes much difference, but one might want to keep an eye on these issues for future reference. The problem is also important from the physics point of view. Having quantified the level at which the matter composition enters the discussion we can compare to (for example) the role of the elastic crust \cite{penner} and  superfluid components \cite{yu}. Finally, and perhaps most importantly, our discussion suggests a ``new'' approach to dynamical tides, putting the mode excitation in focus, not only for resonances associated with the $m\neq0$ modes but also for the $m=0$ contribution. This in interesting as it may allow us to develop new phenomenological models for the all-important gravitational-wave phasing (as an alternative to the effective Love number prescription from \cite{kleff,stein}, see also the recent effort in \cite{schmidt}). 
A first step in this direction is outlined in \cite{nap}.

\end{document}